\documentclass[sigplan,screen,nonacm]{acmart} 

\usepackage{csquotes}
\usepackage{algorithmic}
\usepackage{graphicx}
\usepackage{textcomp}
\usepackage{xcolor}
\usepackage{listings}
\usepackage{tipa}
\usepackage{footmisc}
\usepackage{lineno}               
\usepackage{pstricks}                 
\usepackage{flushend}

\definecolor{codegreen}{rgb}{0,0.6,0}
\definecolor{codegray}{rgb}{0.5,0.5,0.5}
\definecolor{codepurple}{rgb}{0.24,0,0.41}
\definecolor{backcolour}{rgb}{0.96,0.96,0.96}

\lstdefinestyle{cstyle}{
    language=C,
    frame=lines,
    commentstyle=\color{codegreen},
    keywordstyle=\color{codepurple},
    numberstyle=\tiny\color{codegray},
    stringstyle=\color{magenta},
    basicstyle=\ttfamily\footnotesize,
    breakatwhitespace=false,         
    breaklines=true,                 
    captionpos=b,                    
    keepspaces=true,                 
    numbers=left,                    
    numbersep=5pt,                  
    showspaces=false,                
    showstringspaces=false,
    showtabs=false,                  
    tabsize=2
}

\lstdefinestyle{pystyle}{
    language=Python,
    frame=lines,
    commentstyle=\color{codegreen},
    keywordstyle=\color{codepurple},
    numberstyle=\tiny\color{codegray},
    stringstyle=\color{magenta},
    basicstyle=\ttfamily\footnotesize,
    breakatwhitespace=false,         
    breaklines=true,                 
    captionpos=b,                    
    keepspaces=true,                 
    numbers=left,                    
    numbersep=5pt,                  
    showspaces=false,                
    showstringspaces=false,
    showtabs=false,                  
    tabsize=2
}

\begin{document}

\title{Compact Native Code Generation for Dynamic Languages on Micro-core Architectures}
\setcopyright{none}
\author{Maurice Jamieson}
\orcid{0000-0003-1626-4871}
\email{maurice.jamieson@ed.ac.uk}
\affiliation{%
  \institution{EPCC at the University of Edinburgh}
  \streetaddress{Bayes Centre, 47 Potterrow}
  \city{Edinburgh}
  \country{United Kingdom}
}

\author{Nick Brown}
\orcid{0000-0003-2925-7275}
\affiliation{%
  \institution{EPCC at the University of Edinburgh}
  \streetaddress{Bayes Centre, 47 Potterrow, Edinburgh}
  \city{Edinburgh}
  \country{United Kingdom}
}

 \renewcommand{\shortauthors}{Maurice Jamieson and Nick Brown}

\begin{abstract}
Micro-core architectures combine many simple, low memory, low power-consuming CPU cores onto a single chip. Potentially providing significant performance and low power consumption, this technology is not only of great interest in embedded, edge, and IoT uses, but also potentially as accelerators for data-center workloads. Due to the restricted nature of such CPUs, these architectures have traditionally been challenging to program, not least due to the very constrained amounts of memory (often around 32KB) and idiosyncrasies of the technology. However, more recently, dynamic languages such as Python have been ported to a number of micro-cores, but these are often delivered as interpreters which have an associated performance limitation.

Targeting the four objectives of performance, unlimited code-size, portability between architectures, and maintaining the programmer productivity benefits of dynamic languages, the limited memory available means that classic techniques employed by dynamic language compilers, such as just-in-time (JIT), are simply not feasible. In this paper we describe the construction of a compilation approach for dynamic languages on micro-core architectures which aims to meet these four objectives, and use Python as a vehicle for exploring the application of this in replacing the existing micro-core interpreter. Our experiments focus on the metrics of performance, architecture portability, minimum memory size, and programmer productivity, comparing our approach against that of writing native C code. The outcome of this work is the identification of a series of techniques that are not only suitable for compiling Python code, but also applicable to a wide variety of dynamic languages on micro-cores.
\end{abstract}




\maketitle

\section{Introduction}
Micro-core architectures combine many simple, low power, CPU cores on a single processor package. Providing significant parallelism and performance at low power consumption, these architectures are not only of great interest in embedded, edge, and IoT applications, but also demonstrate potential in high performance workloads as many-core accelerators.

From the Epiphany \cite{noauthor_epiphany_2014}, to the PicoRV-32 \cite{wolf_picorv32:_2018}, to the Pezy-SC2, this class of architecture represents a diverse set of technologies. However, the trait that all these share is that they typically provide a very constrained programming environment and complex, manually programmed, memory hierarchies. In short, writing code for these architectures, typically in C with some bespoke support libraries, is time consuming and requires expertise. There are numerous reasons for this, for instance, not only must the programmer handle idiosyncrasies of the architectures themselves but also, for good performance, contend with fitting their data and code into the tiny amount (typically 32 - 64 KB) of on-core fast scratchpad memory.

Little wonder then that there have been a number of attempts to provide programming technologies that increase the abstraction level for micro-cores. Some early successes involved a simple port of OpenCL \cite{stone2010opencl} and OpenMP \cite{openmp}, however, whilst this helped with the marshalling of overall control, the programmer still had to write low-level C code and handle many of the architectural complexities. More recently, a number of dynamic languages have been ported to these very constrained architectures. From an implementation of a Python interpreter \cite{brown_epython_2016}, to \cite{noauthor_parallellaotp_2020} and \cite{ritz_drewvidparallella-lisp_2020}, these are typically provided as interpreters but the severe memory limits of these cores is a major limiting factor to these approaches.

It was our belief that a compilation, rather than interpreted, approach to these dynamic languages would offer a much better solution on the micro-cores. However, the very nature of the technology requires considerable thought about how best to construct an approach which can deliver the following objectives:

\begin{itemize}
\item{Performance at or close to directly written C code}
\item{The ability to handle code sizes of any arbitrary length}
\item{Portability between different micro-core architectures}
\item{Maintaining the programmer productivity benefits of dynamic languages}
\end{itemize}

In this paper we describe our approach to the construction of a compilation approach for dynamic languages of micro-core architectures. The paper is organised as follows; in Section \ref{background} we explore background and related work from a compilation perspective and why existing approaches did not match the objectives described above. Section \ref{codegen} then describes our approach in detail, before applying this to an existing Python interpreter, called ePython, for micro-cores in Section \ref{epython}. In Section \ref{results} we not only explore the application of our approach to compiling Python for micro-cores, but also the performance and memory characteristics of our approach compared to writing directly in C. Section \ref{conclusions} then draws a number of conclusions and discusses further work. 

\section{Background and Related Work}\label{background}
Whilst implementations of dynamic languages, such as Python, for micro-core architectures greatly reduces the time and effort to develop applications in comparison to writing them using the provided C software development kits (SDKs), there remains the performance overhead of interpreting dynamic languages over compiled C binaries. There are two major approaches to accelerating Python codes, just-in-time (JIT) compilation of the bytecodes at runtime and ahead-of-time (AOT) compilation before the code executes. In desktop and server environments, the Numba JIT compiler \cite{lam_numba_2015} accelerates Python applications by specifying a subset of the language that is compiled to native code for central processing units (CPUs) and and graphics processing units (GPUs). Numba uses a Python \emph{function decorator}, or directive, approach to annotating the code to compile to native code, defining the \lstinline[style=pystyle]{@jit} decorator to target CPUs, and the  \lstinline[style=pystyle]{@cuda.jit} decorator to target GPUs and perform the necessary data transfer. Numba's JIT compiler can speed up codes by a factor of around 20 times that of the standard Python interpreter \cite{seif_heres_2019}. 

The Nuitka \cite{hayen_nuitka_2012} and Cython \cite{behnel_cython_2011} AOT Python compilers generate C source code that is compiled and linked against the CPython runtime libraries. Whilst, like Numba, they are highly compliant with, and produce binaries that are significantly faster than, the standard CPython interpreter, up to 30 times faster in the case of Cython \cite{seif_use_2019}, the overall binary size is large. For instance, if we take the seven line code example by \cite{brunet_elf_2020} and compile it on x86 Linux, Nuitka produces a dynamically-linked binary that is 154KB, and Cython produces a much smaller binary at 43KB. However, when the required dynamic libraries are included, the overall size is significantly larger, at 29MB for the archive file produced by Nuitka using the \lstinline[style=cstyle]{--standalone} compiler option. A smaller hand-crafted static binary was produced by \cite{brunet_elf_2020} using Cython but it is still 3MB in size. The binaries generated by Numba, Nuitka and Cython are much larger than the memory available on micro-core architectures, thereby requiring a different approach.

In embedded environments, MicroPython \cite{noauthor_micropython_nodate} uses AOT compilation to accelerate codes for the target microcontrollers. MicroPython implements two native code \emph{emitters}, \emph{native} and \emph{viper} \cite{george_update_4}, defining the \lstinline[style=pystyle]{@micropython.native} and \lstinline[style=pystyle]{@micropython.viper} decorators to select the emitter for native code generation. The \emph{native} emitter replaces the MicroPython bytecode and virtual machine (VM) with machine code representations of the bytecode and calls to VM functions for operations such as arithmetic calculations, binary operations and comparisons. Effectively, this method removes the overhead of the VM's dispatch loop, whilst leveraging the existing VM functions and capabilities. The \emph{viper} emitter takes this approach further by also generating machine code instructions for operations, rather than calling the VM functions. As we would expect, this increases performance yet further as arithmetic operations are performed inline rather than via a procedure call. This results in performance approximately 10 times that of the \emph{native} emitter and around 24 times faster than the VM \cite{george_update_5}. Whilst the MicroPython emitters are not JIT compilers, as the native code is generated before execution begins and the bytecode is not profiled to select compilation candidates, the code generation is performed on the microcontrollers themselves. 

Our native code generation approach is similar to that of the MicroPython \emph{viper} emitter in that native code is generated for all Python code, including operations such as arithmetic and comparisons. However, as will be discussed in Section \ref{codegen}, we generate C source code that is compiled to a native binary for download and execution on the micro-core device. As well as the Nuitka and Cython AOT Python compilers, a number of existing programming languages have used this approach, including Eiffel \cite{eiffel_two-minute_2020}, Haskell \cite{terei_llvm_2010} and LOLCODE \cite{richie_i_2017}. Whilst Haskell has deprecated the C backend in preference to one based on LLVM \cite{noauthor_llvm_nodate}, the former is still beneficial for porting to a new platform as it produces \emph{vanilla code}, requiring only \lstinline[style=cstyle]{gcc}, \lstinline[style=cstyle]{as} and \lstinline[style=cstyle]{ld} tools \cite{noauthor_411_nodate}. Likewise, C was chosen as the backend for ePython native code generation to enhance portability, particularly as a number of the target micro-core architectures, including the Adapteva Epiphany and Xilinx MicroBlaze, are not supported by LLVM. Furthermore, code generators, such as MicroPython's emitters, need to be specifically written to support new processor instruction set architectures (ISAs), and the C backend ensures that native code generation is immediately available on all micro-core platforms that ePython supports. Crucially, our approach generates high-level C source code that retains the overall structure of the Python source program rather than emitting machine code representations of the bytecode (MicroPython) or disassembling / translating the bytecode to native code (Numba). This is in order to leverage the extensive optimisation capabilities within modern C compilers, including register allocation, data flow analysis, instruction selection and scheduling, data dependency management and scalar optimisations \cite{cooper_compiler_2012}.

\section{Code Generation}\label{codegen}
Due to the severely limited memory available on micro-core architectures, an AOT compilation approach is favored over JIT. This also enables us to leverage the C compiler'{s} extensive code optimisation routines, at a higher level over a greater amount of source code, resulting in significantly faster code.

Our approach takes the programming language's generated Abstract Syntax Tree (AST) and then traverses this to generate high-level C source code. Whilst this is a common approach to compiling codes, we are not proposing a simple transliteration from the source language to C here but instead the generation of optimal source code that supports the dynamic features of the source language, whilst optimising memory access and arithmetic operations. The target C code is designed around a set of application programming interfaces (APIs) that implement a form of abstract machine for a generic dynamic object-oriented (OO) programming language. 

This approach of generating C code, with associated macros, is similar to the cross-platform, macro-based, code generation approach used by \cite{koch_algol68_1975}. Such an abstract machine can contain powerful programming abstractions, such as first-class and anonymous functions (lambdas), with their respective closure support. 

\subsection{Handling Memory}\label{memorymodel}
\begin{figure}[ht!]
\centering
\includegraphics[width=0.47\textwidth]{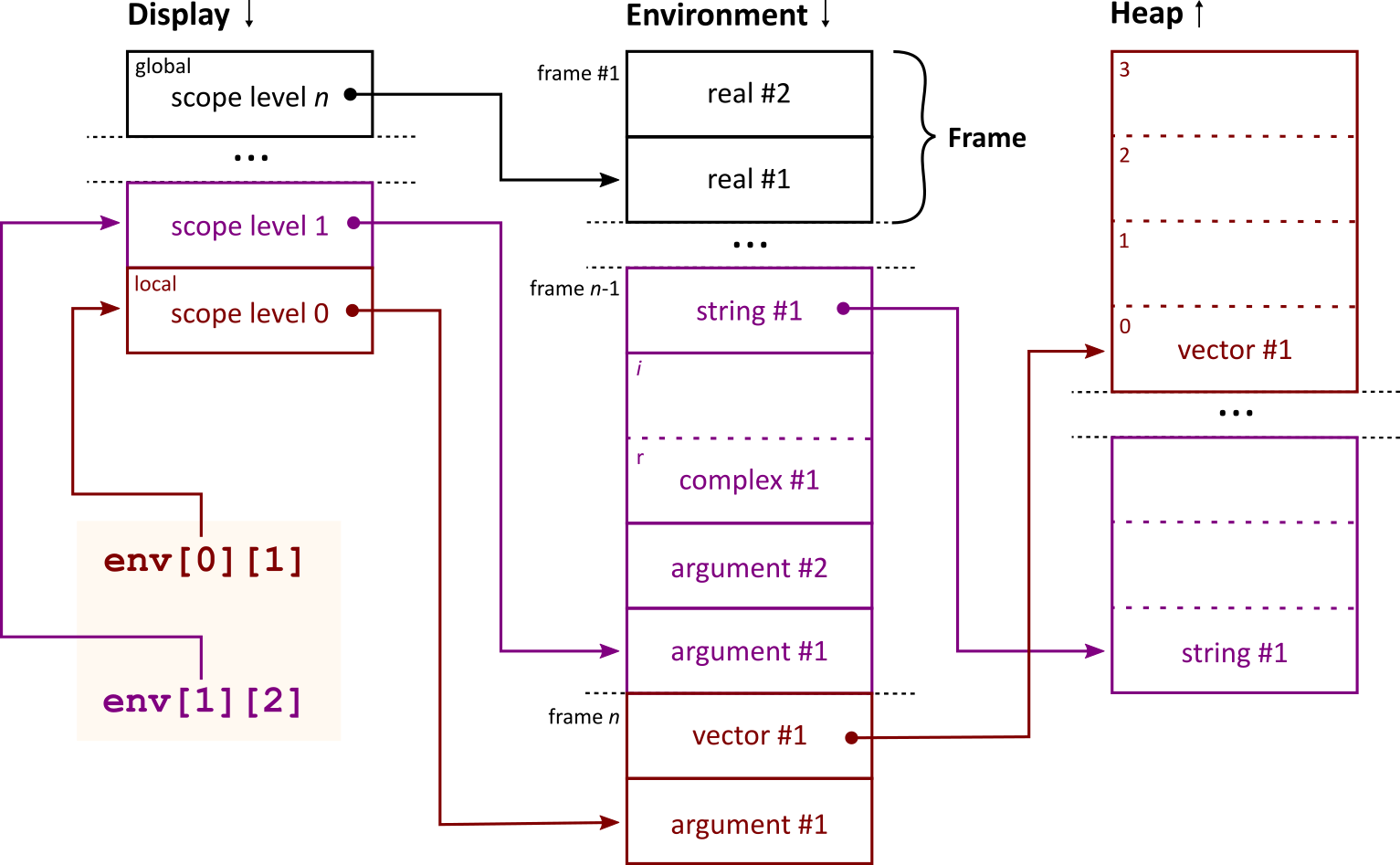}
\caption{Display, environment and frame structure}
\label{fig:frames}
\end{figure}

The main departure of the abstract machine from a simple transliteration is that the target code is not managed using C variables but by the abstract machine through the introduction of environments containing \emph{frames}, as shown in Figure \ref{fig:frames}. Therefore, all memory management, including function stacks and argument passing is managed by the abstract machine, not the C compiler and runtime. Furthermore, as \lstinline[style=cstyle]{libc}, the C runtime library, is often too large to be used successfully on many micro-core architectures, the abstract machine also provides a simple, small heap manager, which can be tailored to the source language in question.

For the sake of simplicity, we will refer to all source programming language elements, for example, numbers, strings, lists, arrays and functions as \emph{variables}. The generated code declares all the variables within frames as shown in Figure \ref{fig:frames}, that grow downwards from the top of memory. Following \cite{hunter_essence_1999}, we create a \emph{display} that holds references to the frames, allowing variable indexing via  \emph{scope level} and \emph{offset}. The compiler calculates and maintains the levels and offsets of the variables within the environment. A new frame is created for each function call and the display is updated to manage the access to variables in outer frames (enclosing scope). For most function calls, a new entry is created in the display, but recursive calls reuse the same display entry updated to point to the new frame for each invocation. This allows us to continue to reference all variables by scope level and offset for recursive functions, thereby maintaining the ability to access all variables by indexing from a base pointer, negating the requirement to chain up the environment list to find variables in outer scope levels. 

Figure \ref{fig:frames} also shows that all fixed-size objects can be allocated in a frame or in the heap. Here, a complex number is declared in frame \#2, rather in the heap as for the vector in frame \emph{n}. Allocating space for variables in the frame is much quicker than allocating them in the heap as the process of memory allocation is much simpler; allocating the space within a frame is just the matter of decrementing the frame pointer by the required amount. This model allows decisions to be made within the compiler regarding the best placement for composite data types.

The display and environment of frames shown in Figure \ref{fig:frames} perform similar capabilities to the C stack. However, for brevity the diagram is simplified, and the environment includes dynamic and static links to support nested functions / closures. A more detailed explanation of these mechanisms can be found in \cite{aho_compilers_1986}, and \cite{bornat_understanding_2020} discusses the issues with simple displays for environment and closure support, outlining the \emph{environment link} mechanism solution, a version of which is used in our approach. Therefore, we will not discuss the detailed implementation of these underlying mechanisms in this paper beyond highlighting that there is a static limit on the maximum number of scope (lexical) levels for kernels running on the device. For a lot of codes, this could be a predefined value but this can also be calculated by the compiler, even for kernels with dynamically loaded functions. For ePython kernels, the compiler \emph{visits} all functions, including those which are dynamically loaded and is able to calculate the required maximum value.

\subsubsection{Variable Access}
Listing \ref{lst:access} is an example of the typed variable access in the generated C code, where the variables are addressed by their scope level and offset, with the memory layout visualised in Figure \ref{fig:frames}. In this example, the \lstinline[style=cstyle]{real} (float) part of the \lstinline[style=cstyle]{complex} variable at the enclosing (non-local) scope level 1 and offset 2 is being set to 4.3. The second line updates the element, indexed by the variable at offset 0, of a vector stored in the heap and declared in the local scope (level=0) at offset 1.

\begin{minipage}{0.95\linewidth}
\vspace{\baselineskip}
\begin{lstlisting}[style=cstyle, caption={Variable access example}, label={lst:access}]
update_complex_real(lookup_complex(env,1,2),4.3);
vector_update_int(lookup_vector(env,0,1),lookup_int(env,0,0),42);
update_real(env,4,1,10.0);
\end{lstlisting}
\end{minipage}

As shown in Figure \ref{fig:frames}, the scope level increases outwards from the local block (or function) scope, with local variables having a scope level of zero. This enables the C compiler to use indexed addressing from the frame pointer (\lstinline[style=cstyle]{env[0]}) to access local variables, thereby increasing performance for local variables and loop block indices, as the compiler can use indexed addressing to directly access the local variables.  Furthermore, the indexing of outer scope levels directly removes the need to chain up environment frames to locate non-local variables, with the corresponding performance overhead. This model also allows support of Python 3 \lstinline[style=cstyle]{nonlocal} variables \cite{noauthor_7_2020} that are declared in the nearest enclosing scope level, as shown in line 1 of Listing \ref{lst:access}. In the target C code, the variable access APIs are implemented as macros that directly update the frame elements within the environment and include the required casting to and from the current variable type. This is illustrated in line 3 of Listing \ref{lst:access}, where the \lstinline[style=cstyle]{update_real} macro expands to \lstinline[style=cstyle]{(((Real*)(env[(lex_level)]))[(offset)]=(Real)((value)))}. This not only accesses the frame element directly but also ensures that the value is stored correctly in memory.

\begin{minipage}{0.95\linewidth}
\vspace{\baselineskip}
\begin{lstlisting}[style=cstyle, caption={Generated function example}, label={lst:function}]
Int oly_e1(Env env, Object self) {
  return(lookup_int(env,0,0) +
         lookup_int(env,0,1));
}
\end{lstlisting}
\end{minipage}

All target C generated functions are passed two arguments, the environment of frames \lstinline[style=cstyle]{env} and \lstinline[style=cstyle]{self}, as shown in Listing \ref{lst:function}. The \lstinline[style=cstyle]{self} argument is a reference to an \lstinline[style=cstyle]{Object} type which enables the abstract machine to support object-orientation, where a function is actually a method of an object. The base, or \emph{native}, functions which are pre-provided in the abstract machine, such as heap management, have the same argument model to allow them to be called and dynamically loaded using the same mechanism as compiler generated functions. The added benefit is that this model also encourages the C compiler to place the function arguments (environment and object) and variable offsets in registers, as much as is possible, on all target platforms. The arguments to a function are declared within the same frame as the function's local variables and form part of the overall frame size. The native code function names are machine generated and prepended with \lstinline[style=cstyle]{oly_} to minimise any name clashes with existing code libraries. As you might expect, and as shown in Listing \ref{lst:declareproc}, functions are also declared within the environment frames. As with other variables, function names (e.g. \lstinline[style=cstyle]{add}) are passed to the abstract machine APIs to enable support for debugging and reflection. 

\begin{minipage}{0.95\linewidth}
\vspace{\baselineskip}
\begin{lstlisting}[style=cstyle, caption={Function declaration example}, label={lst:declareproc}]
declare_proc(env,1,"add",
             mk_proc(oly_e1,env,2));
\end{lstlisting}
\end{minipage}

Crucially we have found that, by leveraging frames within an environment, the decoupling of variables from the underlying C storage mechanisms not only provides support for the dynamic and object-oriented features of many source languages but also for the implementation of dynamically loading of functions which we describe in Section \ref{dynamicloading}. 

\subsection{Reducing the Memory Footprint via Dynamic Loading}\label{footprint}

Within the context of micro-cores and their very limited memory size, a major benefit of our abstract machine approach is that it allows source code functions to be dynamically loaded at declaration or at any later point in the execution of a kernel. Listing \ref{lst:loadproc} demonstrates the simple change to the generated code required to allow the example \lstinline[style=cstyle]{add} function in Listing \ref{lst:declareproc} to be dynamically loaded at declaration. As expected, the code definition in Listing \ref{lst:function} isn't required. The \lstinline[style=cstyle]{load_proc} API call initiates the download of the \lstinline[style=cstyle]{add} function from the host, allocates the space in the device's heap to hold the function code and declares it within the frame. 

\begin{minipage}{0.95\linewidth}
\vspace{\baselineskip}
\begin{lstlisting}[style=cstyle, caption={Dynamically loaded function declaration example}, label={lst:loadproc}]
declare_proc(env,1,"add",
             load_proc("add",env,2));
\end{lstlisting}
\end{minipage}

The final argument of the \lstinline[style=cstyle]{load_proc} API call is the number of arguments in the dynamic function and allows the runtime to create a frame of the correct size. Listing \ref{lst:updateproc} shows how dynamic function loading can be deferred to a later point of execution after declaration. The ability to separate the loading of a dynamic function from the declaration allows the compiler to implement a dynamic code loading strategy tuned to a kernel's particular execution profile. Furthermore, as the code for dynamic functions is stored within the abstract machine heap, it can be discarded (freed) as required, thereby allowing the execution of much larger kernels than is possible with previous static code loading model. Crucially, our environment model automatically enables runtime symbol resolution within the compiled C code, enabling dynamic function loading. 

\begin{minipage}{0.95\linewidth}
\vspace{\baselineskip}
\begin{lstlisting}[style=cstyle, caption={Deferred dynamically loaded function example}, label={lst:updateproc}]
declare_proc(env,1,"add", NULL);
...
update_proc(env,1,load_proc("add",env,2));
\end{lstlisting}
\end{minipage}

\section{Python - a Vehicle for Testing Our Approach}\label{epython}

ePython \cite{brown_epython_2016} is an interpreter which implements a subset of Python and is designed to target micro-core architectures. Designed with portability across these architectures in mind, it has evolved from its initial purpose as an educational language for parallel programming, through its use as research vehicle for understanding how to program micro-core architectures, to supporting real-word applications on the micro-cores. 

As described previously, on-core memory is everything with these micro-cores and whilst previous work around memory hierarchies and remote data \cite{jamieson_brown_epython_2019} allow an unlimited amount of data to be streamed through the micro-core memory, there were still fundamental limits to the code size. This resulted in two major impacts, firstly the size of the Python codes that could be executed on the micro-cores and secondly the number of language features that the ePython interpreter could fully support. 

However, it was our hypothesis that by applying the concepts described in Section \ref{codegen}, then not only would the performance of ePython be significantly improved (a compiled vs interpreted language) but the ability to dynamically load different parts could significantly reduce the memory requirement and enable codes of unlimited size to be executed. Put simply, using the approach described in Section \ref{dynamicloading}, one needs only load at a minimum a resident bootstrapper which contains the core support for marshalling and control of dynamically loaded functions. These can then be retrieved on-demand and garbage collected as memory fills up.

\subsection{Implementing the Dynamic Loader Support}\label{dynamicloading}
\begin{figure}[ht!]
\centering
\includegraphics[width=0.45\textwidth]{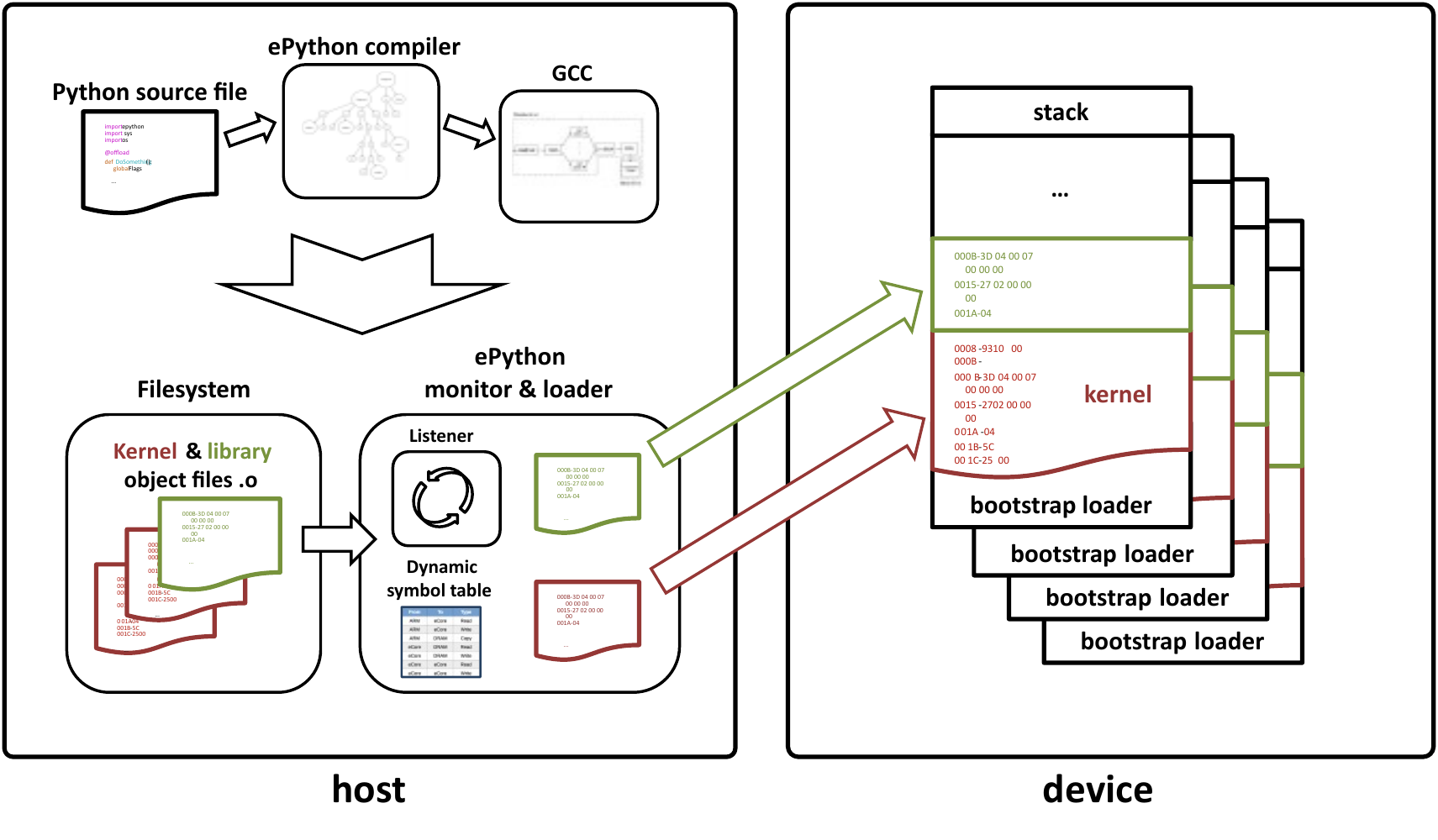}
\caption{ePython dynamic loader architecture}
\label{fig:dynamicloader}
\end{figure}

Figure \ref{fig:dynamicloader} outlines the key components of our dynamic loader when applied to ePython. The updates to support dynamic code loading in the abstract machine were relatively minor; as the memory model already provided the abstraction of the variables and functions from the underlying C runtime, the modifications were mainly concerned with the request / transfer of the dynamic functions from the host and their loading into the heap. However, the changes to the host-based compiler and device support functions were more significant, requiring changes to the compiler, the Python integration module and monitor, and a new object file parser.

There are effectively two options available for dynamic loading; make all user functions dynamic or allow the programmer to select which functions they would like to be dynamically loaded. Initially, we chose the former for ease of implementation but for increased flexibility, ePython now implements the latter by allowing the programmer to annotate dynamic functions, with the key limitation that only top-level functions (kernel entry points) can be marked \lstinline[style=pystyle]{@dynamic}. This does not prevent dynamic functions containing nested functions but these cannot be marked \lstinline[style=pystyle]{@dynamic} individually in order to simplify the management of non-local references. During the traversal of the Python AST during code generation, functions that were annotated with the \lstinline[style=pystyle]{@dynamic} decorator, as shown in Listing \ref{lst:dynamicfunction}, are placed in a separate C source file, with another file containing the bootstrap loader and dynamic loading calls for the relevant functions. A dynamic function symbol table is generated by the compiler and loaded by ePython on the host, which is then used to map the kernel dynamic function requests to the correct object file, highlighted in green in Figure \ref{fig:dynamicloader}. 

\begin{minipage}{0.95\linewidth}
\vspace{\baselineskip}
\begin{lstlisting}[style=pystyle, caption={Function declaration example}, label={lst:dynamicfunction}]
from epython import dynamic

@dynamic(defer=True)
def add(x,y):
  return x+y

@dynamic
def add_nums():
    global add
    add = load_function("add")
    print(add(3,4))
    del(add)
    
add_nums()
\end{lstlisting}
\end{minipage}

\begin{figure}[h!]
\centering
\includegraphics[width=0.40\textwidth]{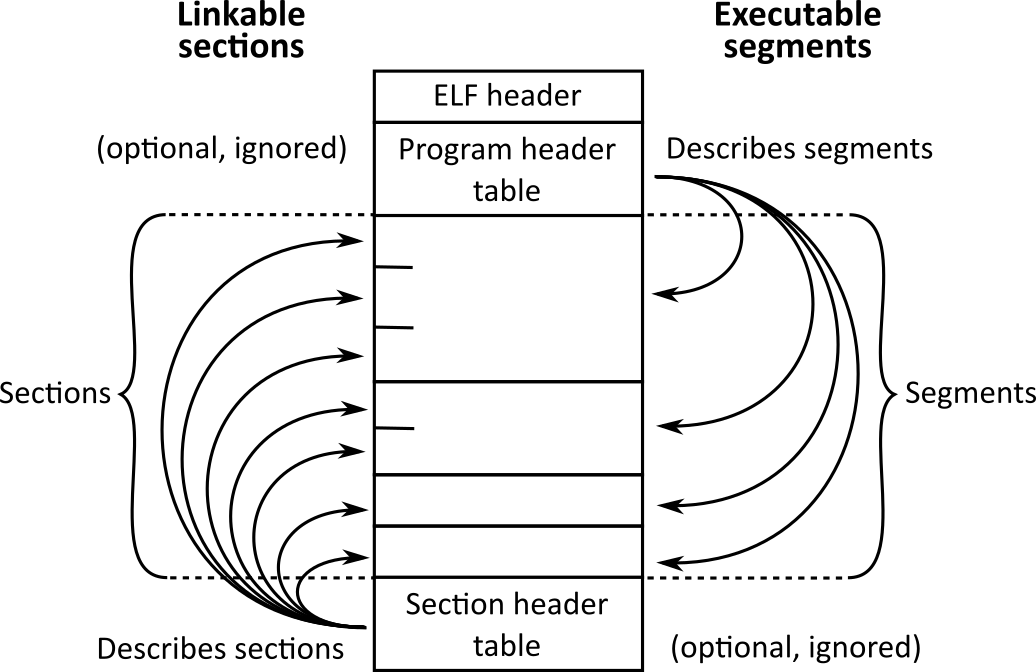}
\caption{ELF file structure \cite{levine_linkers_1999}}
\label{fig:elf}
\end{figure}

The GCC C compiler is used to generate code for the target platforms (Epiphany-III, MicroBlaze and RISC-V, SPARC, MIPS32 and AMD64), resulting in Executable and Linkable Format (ELF) \cite{levine_linkers_1999} object files. Figure \ref{fig:elf} illustrates the ELF file structure, highlighting the linkage between the different sections that need to be traversed to access the required function binary code. The dynamically loaded functions are keyed by their C source name and the host-side symbol table provides the mapping between the ePython source view and generated C function names. 

The dynamic object file created by our approach is parsed when the kernel is downloaded to the devices. The required functions are loaded into memory, based on the entries in the symbol table and then the ELF parser checks that the file is of the correct binary format (for the micro-architecture in question) and raises an error if the file is incorrect for the target device. The ELF object file contains function sizes, which we store in memory along with the functions themselves to allow the device dynamic loader to allocate the correct amount of memory in the heap. As the compiler has previously inserted the number of local variables and arguments into the function declaration code, the abstract machine is able to allocate the correct frame space for future function calls.  

Listing \ref{lst:dynamicfunction} shows how ePython leverages Python's first-class function support to enable the deferment of dynamic function loading, using the \lstinline[style=pystyle]{@dynamic(defer=True)} decorator and mark the function \lstinline[style=pystyle]{add} for deletion from the heap after it has been executed. This allows the programmer to control the exact time during a kernel's execution a function is loaded and marked for deletion. Our model allows the programmer a large amount of flexibility, with the ability to choose whether functions are statically bound to the binary that is downloaded to the device or to download the function at runtime, either when declared or just before execution and retain or delete them from the heap as the kernel execution profile demands.

\subsection{Results and Evaluation}\label{results}
\subsubsection{Code Generation Performance}\label{performance}
\begin{figure}[h!]
	\centering
	\includegraphics[width=0.47\textwidth]{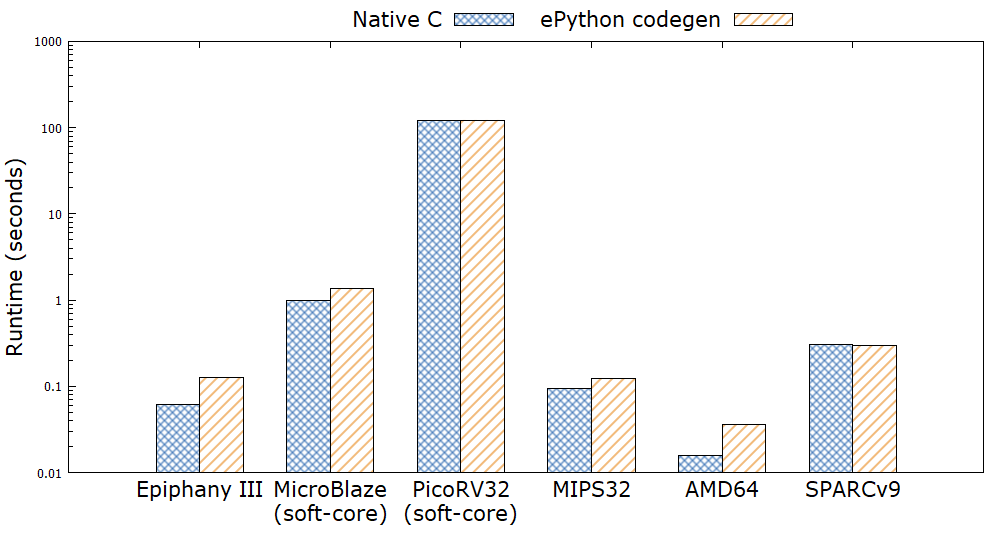}
	\caption{Comparison of native C and ePython codegen performance for the Jacobi benchmark}
	\label{fig:jacobiperf}
\end{figure}

In order to evaluate the level of performance that our code generation model can attain, we used a standard Jacobi code from \cite{noauthor_archer_2016}, that is bundled with the Eithne framework \cite{eithne_poster_2019}. The performance of the native C and ePython codegen versions were compared across a number of CPUs (Epiphany-III, MicroBlaze, PicoRV32, MIPS32, AMD64 and SPARC v9), as shown in Figure \ref{fig:jacobiperf}. The benchmark version for the comparisons was sequential and ran on a single core, with a problem size (NX) of 100 and 10000 maximum iterations on all platforms. Previous experience of generating native code via C for high-level languages suggested that the processor ISA can have a significant impact on both the resulting performance and binary size. Therefore, we included the Intel x86 (AMD64) and SPARC v9 processors as targets. The latter was selected due to its support for \emph{register windows} that, for large C programs, can show a 33\% to 50\% reduction in the number of load and store instructions generated over a non-register window RISC (reduced instruction set computer) processor \cite{catanzaro1991sparc}. Generally, the native C code was faster, as one would expect, but the difference on the soft-cores was surprisingly small. This is likely to be due to the small problem size, where the kernel runtime is impacted by the invocation messaging latency (bandwidth and device listener response). One interesting result is the performance of the Epiphany-III relative to the MIPS32, AMD64 and SPARCv9 processors. This is, in part, explained by the fact that the kernels are executing \emph{bare metal} on the Epiphany-III, MicroBlaze, and PicoRV32, whereas they are running as POSIX threads on the MIPS32 (Linux), AMD64 (Windows Linux Subsystem) and SPARCv9 (Solaris). 

\begin{table}[]
\caption{Relative Jacobi kernel runtime (seconds)}
\centering
\footnotesize
\begin{tabular}{|lll|}
\hline
\textbf{CPU}  & \textbf{Codegen} & \textbf{native C} \\ \hline
\textbf{Epiphany-III}  & 0.12550 & 0.06107 \\
\textbf{MicroBlaze}  &  1.37682  & 0.99044  \\
\textbf{PicoRV32} & 122.80   &  119.84 \\
\textbf{MIPS32}  & 0.12179  & 0.09433  \\
\textbf{AMD64}  &  0.03589 &  0.01563 \\
\textbf{SPARCv9} & 0.30059 &  0.30188  \\ \hline
\end{tabular}
\label{tbl:relativejacobiperf}
\end{table}

Figure \ref{fig:jacobiperf} and Table \ref{tbl:relativejacobiperf} shows that the ePython codegen kernel was slightly faster (0.996 times) than the native C kernel on the SPARC but with the small problem size the runtimes are almost the same for both kernels, suggesting the timings were I/O bound. When the SPARC benchmarks were run again with a much larger problem size (NX=500), the ePython codegen runtimes are 1.140 times slower than the native C version. Comparing this result to the average 33\% performance overhead of the ePython codegen over native C on the other processors suggests our environment model may be leveraging the advantage of the SPARC's register windows. Overall, these are interesting results bearing in mind the runtime's support for the dynamic features of Python versus the static nature of C but it should be noted that these results are \emph{best case} as the compiler is able to generate faster code for this benchmark by removing the need for dynamic function dispatch via environments and making the underlying C function calls directly. We will cover the overhead of the standard dynamic dispatch in Section \ref{dynamicperformance}.

\begin{figure}[h!]
	\centering
	\includegraphics[width=0.47\textwidth]{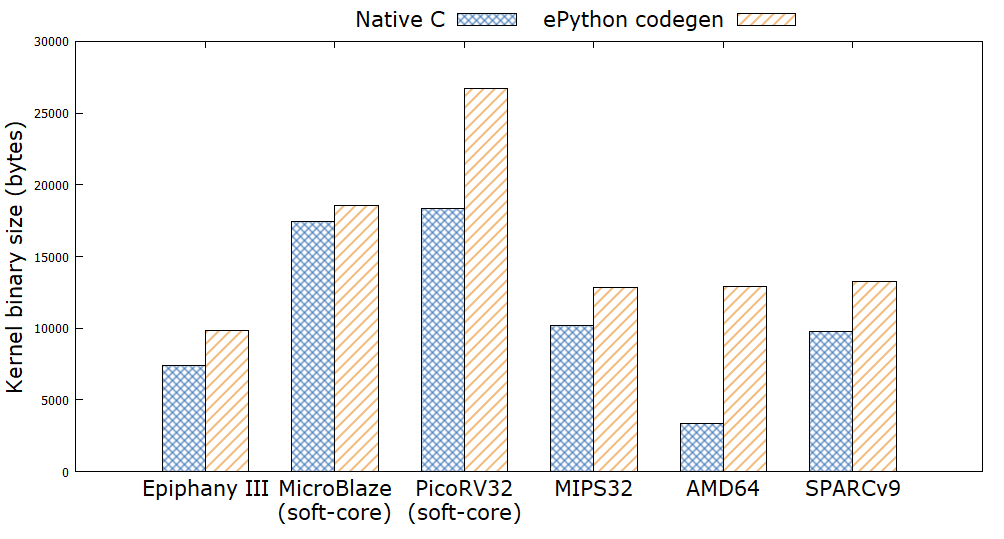}
	\caption{Comparison of native C and ePython codegen kernel size for the Jacobi benchmark}
	\label{fig:jacobisize}
\end{figure}

As on-chip memory is extremely limited on micro-core devices, we compared the relative kernel binary (ELF) sizes for the Jacobi benchmark, as shown in Figure \ref{fig:jacobisize}. However, all the kernels were compiled with the GCC \lstinline[style=cstyle]{-O3} compiler option for maximum performance, rather than \lstinline[style=cstyle]{-Os} for minimum size, as the performance of the offloaded kernels is critical to our target applications and it is crucial to determine if the size of a speed optimised binary would preclude its deployment to the target micro-core architectures.

\subsubsection{Code Generation Code Size}\label{codesize}
As shown in Figure \ref{fig:jacobisize}, the speed optimised codegen binaries are similar in size to native C on the Epiphany-III, MicroBlaze, MIPS32 and SPARC. However, we see a more marked difference on the RISC-V PicoRV32 and the AMD64, suggesting that the C compilers for these processors are more capable at optimising the native C binary for size as well as performance but further investigation of the generated code is required to ascertain if this is the case or if it is a consequence of our environments model on these ISAs. Interestingly, the codegen binary size is within 3\% on the MIPS32, AMD64 and SPARC. From Figure \ref{fig:jacobisize} and Table \ref{tbl:relativejacobisize} we see that the Jacobi codegen binaries on the Epiphany-III, MicroBlaze, PicoRV32 and SPARC are within 46\% of that for native C, with the MIPS32 within 7\% and the AMD64 significantly the worst-case at almost 4 times the size of the native C binary. Bearing in mind that these binaries were compiled for performance (\lstinline[style=cstyle]{-O3}) rather than size (\lstinline[style=cstyle]{-Os}), the overall codegen binary size is viable on the micro-cores as they are all less than 50\% of the available on-chip RAM for the micro-core designs used in the tests. In fact, the Epiphany-III binary only requires approximately a third of the available on-chip 32KB RAM. This significantly enhances the usability of our codegen approach, as the resulting binaries achieve 67\% the performance of native C but still require less than 50\% of the extremely limited memory of the target micro-core devices. These are compelling results bearing in mind the benefits to the programmer: significantly increased productivity and portability across architectures. The challenge with evaluation against other prior work beyond C is that such technologies do not support the tiny memory spaces of our target micro-core devices. Comparable technologies discussed in Section \ref{background}, such as Numba and MicroPython, require far more memory. For example, MicroPython requires at least 256KB, eight times more memory than our target devices. However, we can draw some general comparisons based on previously published work, for instance the MicroPython versus C performance comparison \cite{noauthor_micropython_github_nodate} reveals that MicroPython is approximately 87 times slower than C. If we assume that these values are for the interpreter, and that the ‘Viper’ code generator is 7 times faster \cite{noauthor_micropython_update_2013}, MicroPython is still 12 times slower than hand-crafted C. 

We will provide a comparison of the code generated binary versus the ePython VM, in terms of performance and memory requirements, in Section \ref{dynamiccodesize}.

\begin{table}[]
\caption{Relative Jacobi kernel code size (bytes)}
\centering
\footnotesize
\begin{tabular}{|lll|}
\hline
\textbf{CPU}  & \textbf{Codegen} & \textbf{native C} \\ \hline
\textbf{Epiphany-III}  & 9870 & 7398 \\
\textbf{MicroBlaze}  &  18588  & 17428  \\
\textbf{PicoRV32} & 26728   &  18344 \\
\textbf{MIPS32}  & 12844  & 10160  \\
\textbf{AMD64}  &  12932 &  3350 \\
\textbf{SPARCv9} & 13287 &  9798  \\ \hline
\end{tabular}
\label{tbl:relativejacobisize}
\end{table}

\subsubsection{Dynamic Loader Code Size}\label{dynamiccodesize}
Although, as demonstrated, our codegen approach can generate binaries that are compact enough to run on micro-core architectures, more complex codes still require more memory than is available on the target micro-cores, and this issue is addressed by our dynamic function loading support. As discussed in Section \ref{footprint}, the dynamic loader downloads the required functions from the host and loads them into the abstract machine heap on the device. Since these functions are executed from the heap and not the code segment of the static binary, our method only supports von Neumann and modified Harvard CPU architectures which allow \emph{self-modifying} code. A pure Harvard architecture CPU that has separate code and data buses, with the restriction that code is only executed from the \emph{read-only} code segment is unable to support our dynamic loading model. Therefore, whilst our code generation approach can support generating static binaries on any platform with C99 compiler support, dynamic code loading can only be supported on von Neumann and modified Harvard architectures. With this in mind, we will focus our discussion of dynamic loading on the Adapteva Epiphany-III micro-core, which has 16 von Neumann cores with only 32KB of RAM per core, a restriction that would significantly benefit from being able to dynamically load (and unload) executable code.

In order to understand how different code dispatch and loading models impact performance and code size, the following options were used:

\begin{itemize}
\item{\emph{Static dispatch:} functions are statically bound to the executable and optimised for direct execution}
\item{\emph{Dynamic dispatch:} functions are statically bound to the executable but dynamically dispatched via lookup in the environment}
\item{\emph{Dynamic loading:} functions are dynamically loaded by the kernel and dispatched via lookup in the environment}
\end{itemize}

For the comparisons, we wrote a modified version of the previous Jacobi benchmark to support these options and all ePython codegen and C kernels were compiled with the \lstinline[style=cstyle]{-O3} option, with all dynamic functions compiled with \lstinline[style=cstyle]{-Os}. The latter option ensures the dynamic functions are optimised as far as possible whilst keeping the binary as small as possible. The GCC \lstinline[style=cstyle]{-Os} optimisation option is the same as \lstinline[style=cstyle]{-O2} less any optimisations that increase the size of the code \cite{noauthor_gcc_optimize_nodate}. We also compared the same Python version of the Jacobi benchmark running under the ePython VM. 

\begin{table}[]
\caption{Epiphany-III Jacobi kernel code size (bytes)}
\centering
\footnotesize
\begin{tabular}{|lllll|}
\hline
\textbf{Variant}  & \textbf{Runtime} & \textbf{Bytecode} & \textbf{Functions} & \textbf{Total} \\ \hline
\textbf{ePython VM}       & 21522   & 2329      & -         & 23851 \\
\textbf{Static dispatch}  & 9810    &  -        & -         & 9810  \\
\textbf{Dynamic dispatch} & 12162   &  -        & --        & 12162 \\
\textbf{Dynamic loading}  & 6774    &  -        & 1088      & 7862  \\
\textbf{C}                & 7398    &  -        & -         & 7398  \\ \hline
\end{tabular}
\label{tbl:jacobisize}
\end{table}

Table \ref{tbl:jacobisize} details the overall code size for the different Jacobi benchmark variants, with the VM size including the interpreter, runtime support and bytecode, and the codegen dynamic loading including the static binary (kernel and runtime support), plus the dynamically loaded functions. The non-applicable entries in the table are marked with a dash, for example, bytecode is not applicable for the codegen and native C variants. As expected, the native C variant of the code is the smallest, with the dynamically loaded codegen variant trailing by only 464 bytes. This is followed at 9810 bytes by the codegen variant with all functions statically bound and dispatched, where the compiler is free to optimise function calls. The dynamic dispatch code gen variant is 24\% bigger, as the dynamic dispatch method requires more code (lookups in the environment) and prevents the compiler from optimising the function calls. Finally, the ePython VM variant of the benchmark has the smallest compiled code (bytecode) size but requires the VM of around 22KB to execute. It should be noted that the static dispatch option can only be used for very simple codes, where the compiler can detect simple function calls that do not need new frames, for example, those without local variables or are non-recursive. Therefore, for most functions the compiler will generate code that uses the dynamic dispatch model. Whilst this option produces code that is approximately 64\% larger than native C, it uses only 51\% of the overall memory required by the ePython VM variant. Therefore, for the same Python source code, significantly more of the Epiphany's limited 32KB on-core memory is available for data. 

\begin{figure}[h!]
	\centering
	\includegraphics[width=0.47\textwidth]{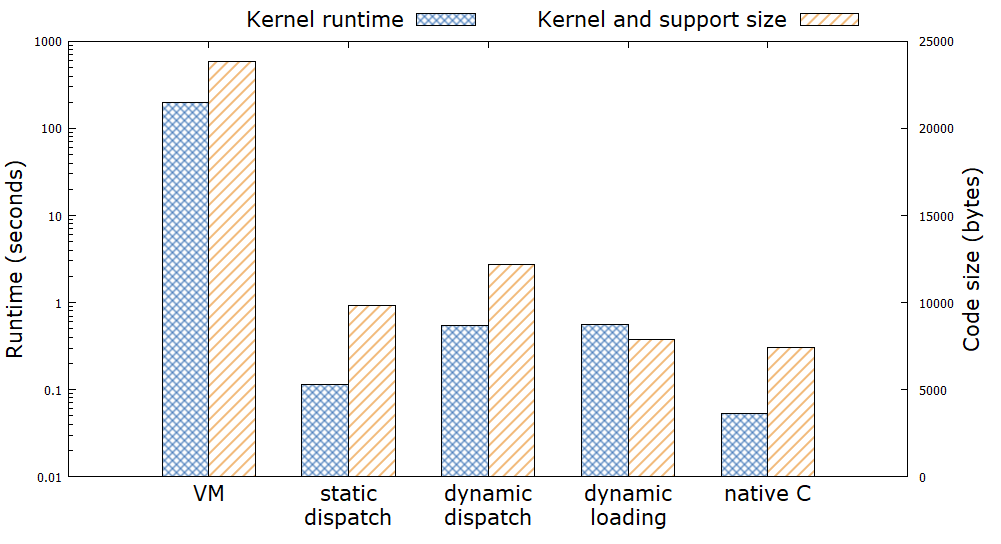}
	\caption{Relative performance and code size for the modified Jacobi benchmark on the Adapteva Epiphany}
	\label{fig:jacobisizeperf}
\end{figure}

\subsubsection{Dynamic Loader Performance}\label{dynamicperformance}
Figure \ref{fig:jacobisizeperf} shows the relative code sizes and runtime performance for all the benchmark variants. Unsurprisingly, the native C benchmark kernel has the fastest execution time at 0.053 seconds and the static dispatch codegen variant is around 2 times slower at 0.115 seconds. The dynamic dispatch model decreases performance relative to the static dispatch model by approximately 5 times at 0.539 seconds and the dynamic loading version is only marginally slower at 0.566 seconds. As we can see from Figure \ref{fig:jacobisizeperf} and Table \ref{tbl:jacobiperf}, the default dynamic function dispatch codegen version is significantly faster than the VM version, which requires around 201 seconds to execute the same Python kernel. However, it should be noted that the ePython is also executing the bytecode from significantly slower off-chip memory (150 MB/s maximum bandwidth obtainable in practice \cite{castro2018energy}), as the kernel bytecode and heap requirements are too large to allow the bytecode to execute from on-chip RAM. When we consider that the dynamic loading version code also only requires approximately a third of the memory to execute, we not only gain a significant performance increase, but also have the ability to handle much more data at the same time, as well as being able to execute arbitrary sized codes via dynamic function loading. This greatly increases the practical applications that dynamic languages can support on micro-core architectures.

\begin{table}[]
\caption{Epiphany-III Jacobi kernel runtime (seconds)}
\centering
\footnotesize
\begin{tabular}{|ll|}
\hline
\textbf{Variant}  & \textbf{Runtime}  \\ \hline
\textbf{ePython VM}       & 201.54   \\
\textbf{Static dispatch}  & 0.11492 \\
\textbf{Dynamic dispatch} & 0.53895   \\
\textbf{Dynamic loading}  & 0.56578  \\
\textbf{C}                & 0.05260   \\ \hline
\end{tabular}
\label{tbl:jacobiperf}
\end{table}

\section{Conclusions and Further Work}\label{conclusions}
The micro-core classification covers a wide variety of processor technologies and this is a thriving area which contains a number of vibrant communities. Whilst these are very interesting for a number of different reasons, a major challenge is around programmer productivity. Whilst we firmly believe that Python has a significant role to play here, the performance impact of a traditional interpreter greatly reduces its viability for high-performance applications on these technologies. In this paper, to address this, we have introduced a code generation approach for dynamic languages, such as Python. Our code generation model was specifically designed to address these performance concerns and to be able to support the peculiarities of micro-core architectures, and more specifically the simplicity of the cores themselves and tiny amounts of associated memory. The reader can clearly see that our approach is widely portable to a number of different processor architectures, whilst also returning small, high performance code that significantly releases precious on-chip memory that can be allocated for data. Furthermore, our dynamic loading model enables further memory savings, as well as supporting arbitrary sized codes, whilst returning performance that, although is approximately 10 times slower than optimised native C code, is over 300 times faster than the interpreter. Crucially, our code generation and dynamic loading approach provides the compiler with options to optimise the resulting code in terms of static function dispatch where possible, dynamic dispatch where required and points for the loading (and unloading) of dynamic functions, based on the execution profile of the code.

Therefore, our present focus is in maturing the native code generation as we think this has demonstrated some worthwhile early results. Further work includes exploring opportunities for further performance improvements, validated by a wider range of benchmarks to provide greater coverage of the abstract machine design. Furthermore, currently the architecture specific runtime library is not included in the dynamic loading. Through extending the dynamic loading approach to include the runtime support, the minimum size will be around 1.5KB plus the size of the largest function. This will open up the possibility of running over a number of additional micro-core architectures that contain tiny amounts of memory per core (less than 8KB). 

The environment model provides other opportunities; functions have a reference to their environment (closure) that can be traversed, and it would be possible to persist functions for later execution, enabling task switching of functions or interrupt support within the abstract machine. The code bodies for these functions could remain in memory for faster switching response or could be unloaded for longer interruptions to free critical memory resources. 

Whilst this paper has focused on a code generation model using Python as a vehicle for testing our approach, we also believe that the work here has a wider applicability to other dynamic programming languages targeting micro-core architectures. 

\pagebreak


\bibliographystyle{ACM-Reference-Format}
\bibliography{./bibliography/IEEEexample}

\end{document}